# Vibrational methods for the symmetrization of asymmetric laminar viscous fluid flow in a plane diffuser

Alexey I. Fedyushkin, Andrei A. Gnevushev, Artem S. Zakharov

*Ishlinsky Institute for Problems in Mechanics of Russian Academy of Sciences Moscow,*
*E-mails: fai@ipmnet.ru, gnevushev-98@mail.ru, zakharovartemm@yandex.ru*

**Abstract**—This paper presents the results of numerical modeling of laminar flows of a viscous incompressible fluid in a flat diffuser with and without vibration effects. Two methods of flow symmetrization in a flat diffuser are considered: by means of periodic vibration action at the inlet to the diffuser and on its walls. Modeling has shown that it is possible to symmetrize the flow of a viscous liquid in the diffuser by using a weak harmonic vibration action, the speed of which can be less than 0.01% of the speed of the main flow.

## 1. INTRODUCTION

The study of viscous fluid flows in a flat diffuser, depending on the intensity, geometric parameters and opening angle is an extensive class of phenomena that occur in nature and are important in various fields of application. In this regard, the study of the regularities of fluid flow in the diffuser is an actual issue.

Many articles have been published on fluid dynamics in flat diffusers. However, despite considerable interest in the study of fluid flows in diffusers, less attention has been paid to the study of stationary asymmetric laminar flows and their symmetrization than to the study of nonstationary and turbulent fluid flows.

The first analytical solutions for describing the symmetric flow of a viscous incompressible fluid flowing from a point source in an infinite diffuser were independently obtained by Jeffery and Hamel more than a hundred years ago [1, 2].

However, the laminar flow of a viscous, incompressible fluid in a flat diffuser retains its symmetry with respect to the center plane only at relatively low values of the Reynolds number. If the critical value of the Reynolds number is exceeded, the flow in the diffuser may become asymmetric. The symmetry of the fluid flow is disrupted due to a local decrease in tangential friction stresses near the walls [3, 4] and, as a result, the appearance of weak secondary vortices near the walls and the separation of the laminar boundary layer from the solid walls. In a flat diffuser, when the critical value of the Reynolds number is exceeded, the separations of the laminar boundary layers on different walls occur at different distances from the inlet. That is, secondary vortices form at the walls of the diffuser, located along the walls in a staggered manner with opposite rotations. The non-uniqueness and bifurcation of solutions to the Jeffery-Hamel problem with small Reynolds numbers have been discussed in papers [5-21]. In [5] a general solution of the Jeffery-Hamel problem in elliptic functions is found and the possibilities of various forms of flows and their stability are discussed. In the paper [6] analyzes the solution to the Jeffery-Hamel problem for a diffuser with slightly curved walls and parabolic velocity profiles. These results suggest the possibility of separation of the laminar flow. The authors of [7] showed the existence of multiple bifurcations that violate the flow symmetry in a finite domain, and that isolated non-unique solutions that are stable in time can exist. These states are analogs of stationary waves in a finite domain, represented in [13] for an infinite domain.

The authors of [8, 9] experimentally investigated the flow in the diffuser and behind the protrusion. In [8], the authors showed the existence of non-stationary flow regimes within the framework of constantly repeating secondary flows near the walls and separation from the walls of the main flow with a "sudden jump" of separation from one wall of the diffuser to another. It was shown in [9] that for small Reynolds numbers in a symmetric channel with a stepwise expansion, the flow can have a stationary asymmetric character.

In [10] presents the results of 2D and 3D (with the symmetry condition) numerical simulation of the flow of a viscous incompressible fluid for Reynolds numbers in the range from 60 to 360 and the angles of the diffuser solution from 10° to 180° in a flat diffuser with the presence of an input section and shows the effect of three-dimensionality and velocity oscillations at the input in the diffuser for the presence of non-stationary asymmetric flow modes.

Reviews of papers on the solution of the Jeffrey-Hamel problem, symmetry breaking, and uniqueness of the solution were made in papers [11 -18]. In [11], based on group analysis, it was also stated that inhomogeneities are possible in stationary solutions to the Jeffery-Hamel problem. That is, there is a possibility of asymmetry and bifurcation of flow in the fluid flow in a diffuser. In [12] for laminar regimes, the results of the existence of stationary asymmetric flow regimes in the diffuser are presented.

In [13] authors studied the occurrence of stationary and non-stationary waves with special symmetry in the flow of a liquid in a diffuser at Reynolds numbers between 5 and 5000, at different angles of diffuser opening. The dependences and asymptotic behavior of wave existence were determined as a function of diffuser angle and Reynolds number, and it was found that these waves were not observed in Poiseuille flows.

In [14], it was shown that finite disturbances at the inlet and outlet of the diffuser at large values of the Reynolds number can violate the Saint-Venant's principle. When there is a stable flow in time and it is unstable in space, then small changes at the inlet or outlet create a significant effect on the flow in the entire channel. However, for small Reynolds numbers, the authors of [14] did not find this effect. In addition, the authors obtained a significantly lower value of the critical Reynolds number when the flow becomes asymmetric than it was in the known experiments and in the article [12].

In papers [15, 16], the results of studying the flow in a finite diffuser based on numerical modeling are presented. These results indicate the peculiarities of the occurrence of bifurcations in a finite diffuser compared to an infinite one. The authors [15] point out that during a bifurcation, the radial behavior of the internal mode that violates symmetry in a finite domain is a superposition of two modes, and not a single mode as in an infinite domain. It was found in [15] that the critical Reynolds number for the flow in the finite region is relatively insensitive to the exact choice of boundary conditions at the inlet/outlet and is especially stable in the limit when the opening angle tends to zero. Based on the global stability analysis performed for spatially evolving flow in a diffuser with constant divergence, the authors [16] conclude that the resulting disturbances are not wave-like in flow direction even at constant Reynolds number. Given previous global studies of stability in boundary layers where there is no qualitative difference between global and parallel modes, this result is unexpected. In fact, although there have been many studies on spatially evolving flows, extended but non-undulating regimes are usually not observed. In [16], it is argued that Robin boundary conditions are not appropriate as is usual in studies of this type.

The authors of [17, 18] obtained solutions to the Jeffery-Hamel problem using an analytical-numerical method for solving nonlinear differential equations and pointed out the existence of unique stationary flows in a diffuser. In [17,18], critical values for the opening angles of the diffusers and the Reynolds numbers of transition from single-mode to double- and triple-mode flow were found, indicating the presence of asymmetric solutions for certain ranges of Reynolds numbers. It was also shown that the transition to asymmetric flow occurs at a $Re^* = 18.8/\alpha$, where $\alpha = \beta\pi/180$ is the opening angle in radians. For an opening angle of 4 degrees, the critical value is Re = 269, consistent with the results [12]. In [17, 18] also showed the dependence of the occurrence of bifurcations on the opening angle of the diffuser and found that the bifurcation flow pattern does not change at small scattering angles of the solution (less than 10 degrees).

Stationary laminar flows in a diffuser of non-Newtonian liquids were numerically studied in the article [19]. In [19] present a studying of the influence of the type of velocity profile shapes set at the inlet to the diffuser (constant, Poiseuille, symmetrical and asymmetric) on the flow velocity in the diffuser is also given, which showed the restoration of asymmetric profiles inside the diffuser to the same, under the condition of constant mass flow.

The studies of influence of vibrations on the fluid flow in the diffuser are presented in the papers [20, 21].

## 2. PROBLEM STATEMENT AND MATHEMATICAL MODEL

The laminar flow of a viscous incompressible liquid between two flat walls diverging from each other at a small angle is considered, that is, the flow in a flat finite diffuser of large elongation (Fig. 1).

Figure 1(a, b) shows geometry and example mesh of flat diffuser model. The zones of the inlet and outlet parts of the diffuser on an enlarged scale are shown in Fig.1 (b). The input and output boundaries have the shape of arcs $l_{in}$ and $l_{out}$ which are located from the origin of the coordinate system at the $r_{in}$ and $r_{out}$, respectively.

The purpose of this numerical simulation is to study the possibility of symmetrization of an asymmetric flow under the influence of harmonic vibrations on the flow in diffuser. The paper considers two ways of vibration action: 1) the effect on the inlet velocity of the diffuser and 2) the effect on flow from the walls of the diffuser.

In diffusers with opening angles less than 10 degrees ($\beta < 10°$), the bifurcated flow pattern is almost independent of the opening diffuser angle [17, 18]. It is also known that for small angles the pressure recovery coefficient decreases and is minimal at an angle of about 6 degrees. This paper presents a numerical study of viscous, incompressible flow in a flat diffuser with a small opening angle of 4 degrees and various Reynolds numbers under laminar asymmetric conditions. Flows in diffusers were considered with Reynolds numbers Re ranging from 0 to 560 and vibration forces corresponding to vibration Reynolds numbers $Re_{vibr}$ between 0 and 240.

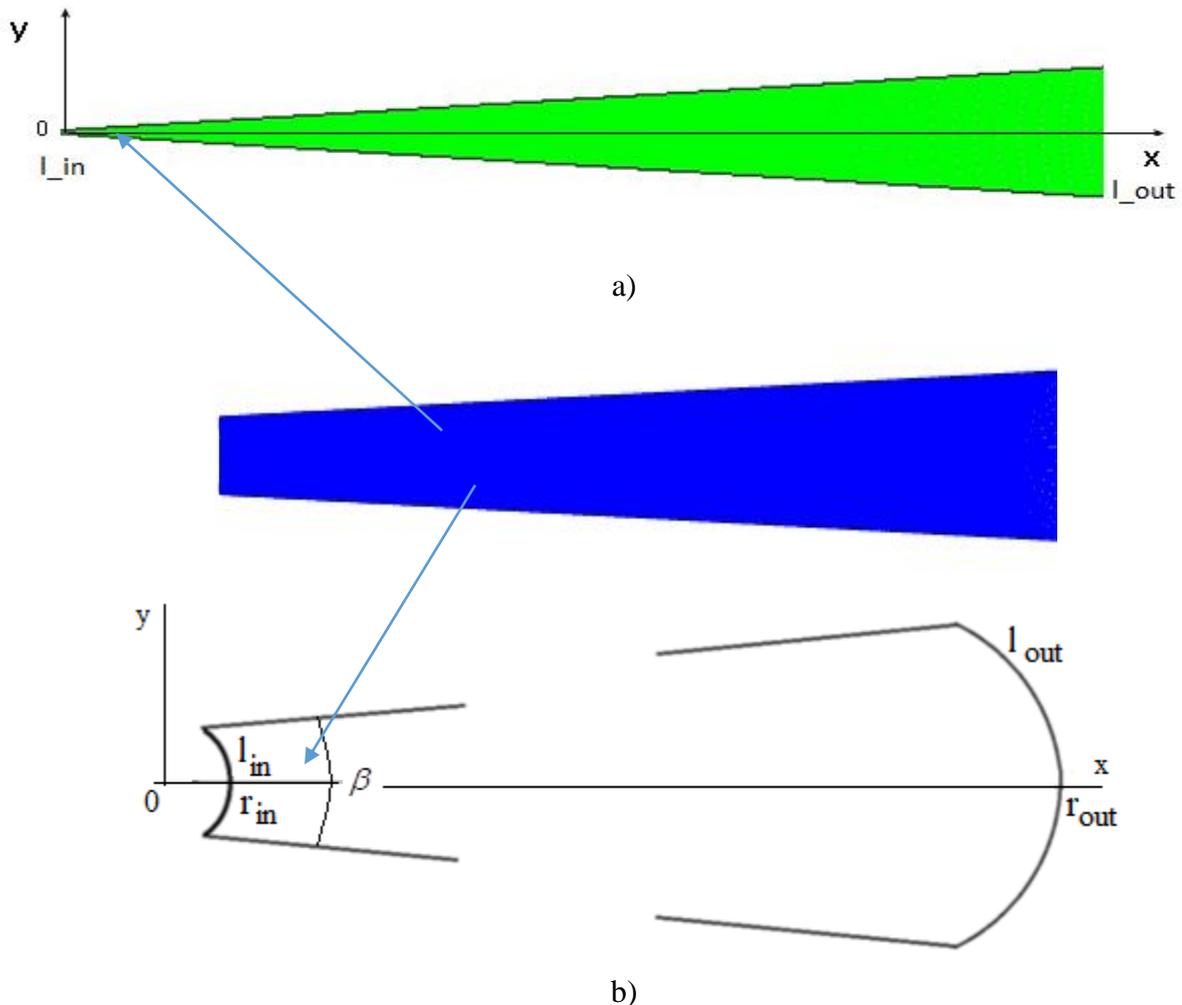

Fig. 1. Schema of calculation domain of a flat diffuser model with an opening angle $\beta$ and length $r_{out}/r_{in} = 100$, a) a general view of the design area and a grid, b) enlarged zones of the input and output zones of the diffuser.

The geometric model of the diffuser is as follows: the opening angle is $\beta = 4°$, the input boundary has the shape of an arc $l_{in}$, located from point the origin coordinates at distance $r_{in} = 0.005$ m, and the output boundary has the shape of an arc $l_{out}$, located from point the origin coordinates at a distance $r_{out} = 0.5$ m (figures 1). Thus, the length of the diffuser L is $L = r_{out} - r_{in} = 0.495$ m (where $r^2 = x^2 + y^2$). This elongated geometry eliminates the influence of boundary conditions at the inlet and outlet and allows you to track the flow change along the length of the diffuser. In particular, it can be observed how quickly the profile returns to a normal value at a constant inlet velocity, corresponding to the Reynolds number and the adhesion conditions on the walls (in [15] that boundary conditions were called "artificial conditions" and it was noted that they do not affect the bifurcation pattern of the flow at small opening angles of the diffuser and low Reynolds numbers).

The problem is simulated based on the numerical solution of a system of two-dimensional nonstationary Navier equations – Stokes for an incompressible viscous fluid for conditions without gravity (1).

$$\frac{\partial V}{\partial t} + (V\nabla)V = -\frac{\nabla P}{\rho} + \nu\Delta V$$

$$\text{div} V = 0 \tag{1}$$

where $V$ ($V_x$, $V_y$) is the velocity vector; $P$ is pressure; $\rho$ is density; $\nu$ is the coefficient of kinematic viscosity, $t$ is time.

The boundary conditions were as follows: at the entrance to the diffuser $l_{in}$, a constant velocity $V_{in}$ was set, directed normally to the boundary $l_{in}$, or in the case of vibrational effect on the inlet boundary in the form $V_{r=r_{in}} = V_{in} + A\sin(2\pi ft)$, where $A$ and $f$ are the amplitude and frequency of the vibrational effect. Pressure is set at the outlet $l_{out}$, a coupling condition is set for velocity at the upper and lower walls of the diffuser $V = 0$ or in the case of vibration from the walls of the diffuser, a condition is set for the walls $V_n = A\sin(2\pi ft)$ (where $V_n$ is the velocity normally directed to the wall). The initial velocity and pressure values were set to zero. The numerical solutions were analyzed in the stationary or quasi-stationary mode of stationary flow oscillations. The Reynolds number at the entrance to the diffuser is defined as $\text{Re} = V_{in}l_{in}/\nu$, where $l_{in}$ is the arc length at the entrance to the diffuser. The Reynolds oscillatory number is defined as $\text{Re}_{vibr} = Al_{in}/\nu$. The vibrational effect on flow in the diffuser depends not only on amplitude, but also on frequency of this effect. Therefore, characteristic dimensionless number can be the vibrational Reynolds number written in form: $\text{Re}_{vibr}^f = A^2/f\nu = \text{Re}_{vibr}/\text{Sh}$, where Sh is the Strouhal number.

Numerical solution of the two-dimensional Navier-Stokes equations (1) was performed using the control volume method and the finite difference method [25-27], which were previously tested on various tasks [24, 26, 27]. In the spatial approximation of the Navier- Stokes equations were solved using numerical methods of second and third orders of accuracy, such as QUICK and MUSCL. These methods are implemented by adding expressions for flow limiters to the right-hand side of the equations, like in [24]. An implicit joint (matrix) method was used for solving equations with greater accuracy compared to the method of separate solution of equations. A conjugate gradient method was also used to solve a system of linear algebraic equations.

For approximating time derivatives, both first- and second-order precision schemes were used. The geometric diversity of the computational domain of the considered diffuser model ($r_{out}/r_{in} = 100$, $L/l_{in} = 1414$, $L/l_{out} = 14.14$) imposes additional requirements on the step size of the mesh in order to eliminate approximation errors. Fig. 1(a) shows a scheme of the calculation area with one of the grids used in the calculations (uneven with rectangular cells near the boundaries), with more than $10^6$ cells. The mesh spacing decreases towards the borders of the diffuser so that there are 5-10 cells in boundary layers. Calculations were carried out on different meshes, and the convergence of numerical results was checked depending on grid step.

In this paper, the following notation is used for dimensionless quantities: $y\_\text{dimless} = y/r\sin(\beta/2)$, $x\_\text{dimless} = x/r$, $V_{x\_\text{dimless}} = V_x/V_{in}$, $V_{y\_\text{dimless}} = V_y/V_{in}$, where $V_{in}$ is the velocity at the entrance to the diffuser. As a criterion for asymmetry, the criterion of changing the angle of inclination of the velocity vector $\phi = \text{arctg}(V_x/V_y)$ at each point in the calculated domains on a steady (or quasi-steady) flow regime is used.

## 3. RESULTS OF THE NUMERICAL SIMULATION

The results of numerical simulation are presented for the flow of a viscous incompressible fluid through a finite diffuser ($r_{in}/r_{out} = 100$, $L/l_{in} = 1414$, $L/l_{out} = 14.14$) with an opening angle $\beta = 4^0$ (Fig. 1) for symmetrical and asymmetrical flow modes, within the range of Reynolds numbers $0 < Re < 560$ without influence of vibrations and with action of harmonic vibrations from the walls of the diffuser in the form $V_n = A\sin(2\pi ft)$ or from the entrance of the diffuser in the form $V = V_{inlet} + A\sin(2\pi ft)$ (for amplitudes $10^{-4} \text{ m/s} < A < 100 \text{ m/s}$ and frequencies $10^{-2} \text{ Hz} < f < 10^2 \text{ Hz}$, which corresponds to $2 \cdot 10^{-3} < Re_{vibr} < 240$ and Strouhal numbers $3.5\,10^5 < Sh < 3.5\,10^2$). In the article, all the results are presented for stationary or quasi-steady state time modes.

### 3.1 The diffuser fluid flow without vibration

*Re=250*

It is known that the flow in a flat diffuser with a small opening angle $\alpha$ at Reynolds number $Re < Re_c = 18.8/\alpha$ is symmetric and stationary [17-19]. Such a flow at Reynolds number $Re = 250$ is presented in Fig. 2. Figure 2 shows velocity isolines (a), isolines of the angles of velocity vectors $\phi = \text{arctg}(V_x/V_y)$ (b), velocity profiles (c-d) and an angle $\phi$ (e) in four vertical sections: $x_{dimless} = 20, 40, 60, 80$.

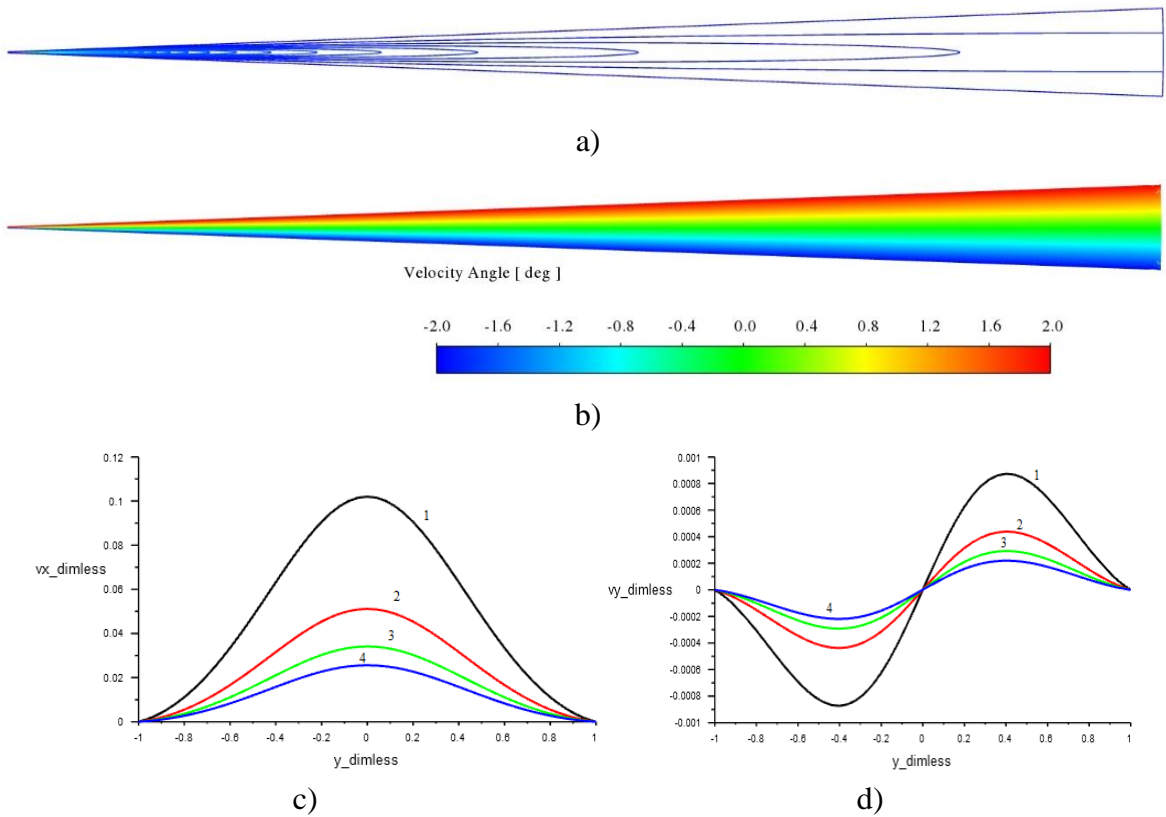

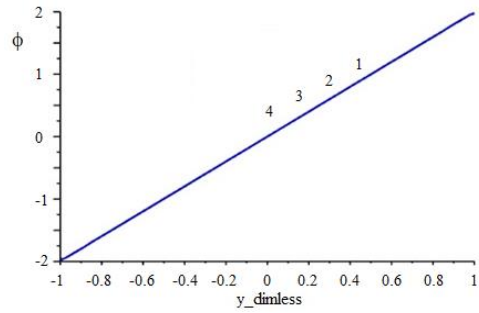

e)

Fig. 2. Velocity isolines (a), isolines of the angles $\phi$ of velocity vectors (b), profiles of dimensionless velocity components (c-d) and the profiles of angles $\phi$ (e) in four vertical sections: (lines 1-4 correspond to sections: $x_{dimless} = 20, 40, 60, 80$) with the Reynolds number Re=250.

In Fig. 2 the profiles of velocity components and the profiles of the angles $\phi$ for Re = 250 are shown. The profiles of angles for the vertical sections $x_{dimless} = 20, 40, 60, 80$ in dimensionless coordinates are the same. The calculations show that the velocities, isolines, and profiles of angles $\phi$ in dimensionless coordinates for all Reynolds numbers below the critical value are identical to those in Fig.2.

*Re=279*

In papers [18, 19], it was found that, in a flat diffuser with an opening angle 4 degrees, the flow becomes asymmetric relative to the median longitudinal plane when Reynolds number (Re) exceeds of the critical value $Re_c$ = 269. An example of asymmetric flow of viscous liquid through a flat diffusor in stationary mode (without vibration) at Re=279 is presented in figure 3. In Fig. 3 the results of calculations for asymmetric stationary flows in a diffuser with Reynolds number Re=279. (a) isolines of the horizontal velocity component Vx, (b) isolines of angle $\phi$ of velocity vectors, (c) profiles of velocity components Vx and Vy, and (d) profiles of the angles $\phi$ in vertical sections (lines 1-4 correspond to different sections $x_{dimless} = 20, 40, 60, 80$) are shown.

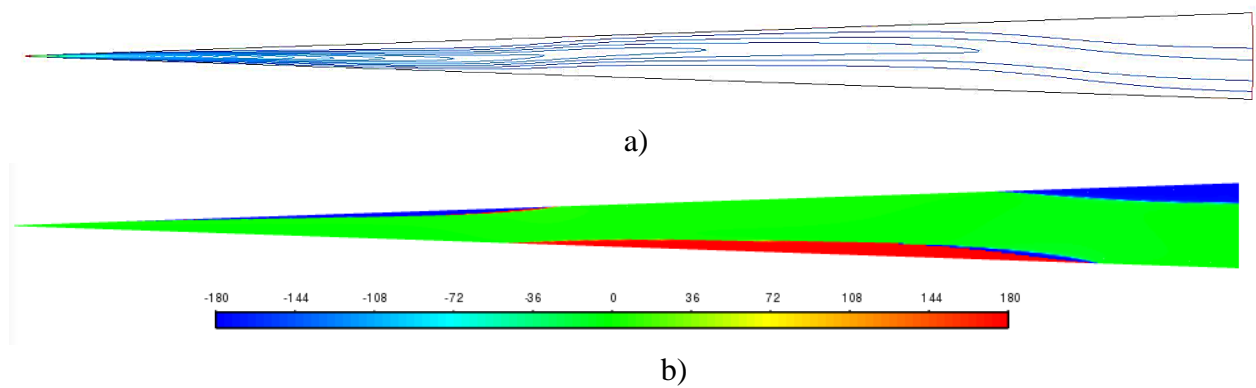

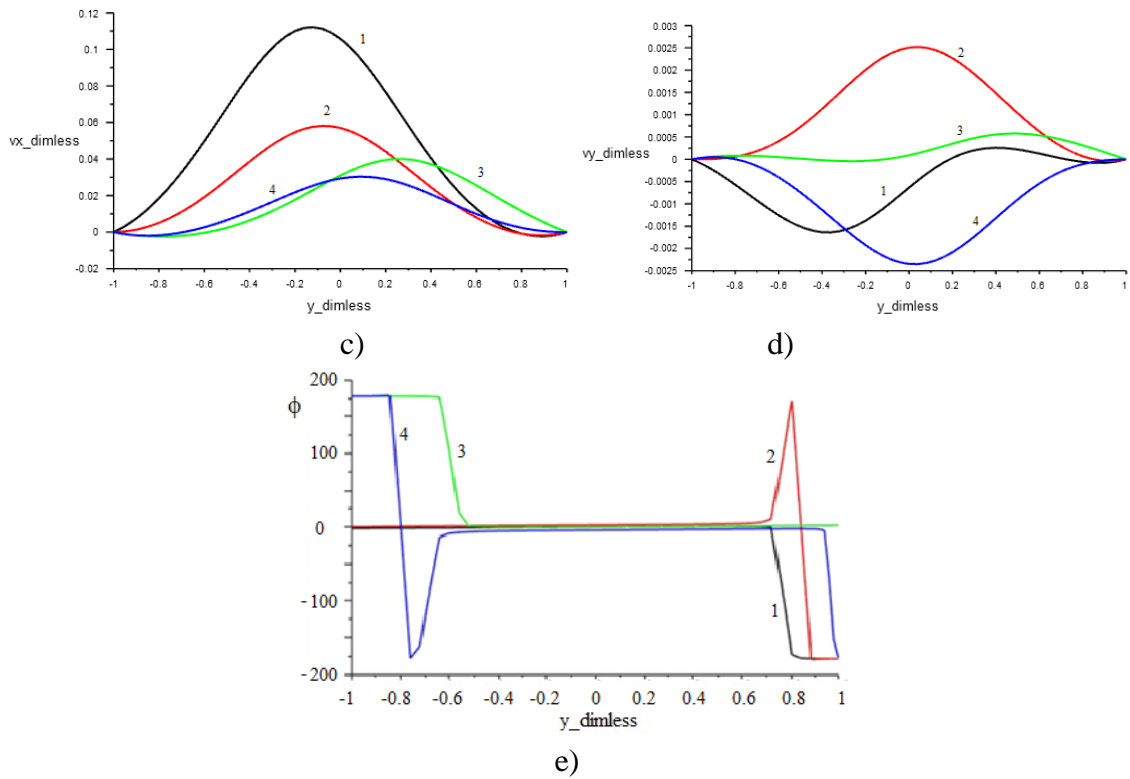

c)

d)

e)

Fig. 3. Velocity isolines (a), isolines of the angles $\phi$ of velocity vectors (b), profiles of dimensionless velocity components (c-d) and profiles of the angles $\phi$ (e) in four vertical sections (lines 1-4 correspond to cross sections with $x_{dimless} = 20, 40, 60, 80$) at Reynolds number Re=279.

*Re=559*

Figure 4 shows the results of numerical simulation of asymmetric flow in a diffuser at Re = 559. For a given Reynolds number, the flow pattern in the diffuser assumes a non-stationary character, and the flow is interspersed between the walls of the diffuser. That is, its appearance and characteristics change both in space and in time. Figure 4 shows: a) – isolines of the horizontal component of the velocity vector Vx; b) – isolines of the angle of direction of the velocity vector , c) and d) –profiles of the components of the dimensionless velocity of instantaneous Vx and time–averaged mean_Vx, respectively; e) - profiles of the angles $\phi$ in vertical sections (lines 1-4 in the figures correspond to sections at $x_{dimless} = 20, 40, 60, 80$). In steady-state mode, velocity values fluctuate over time at each point in the calculated area around their average values. The mean square deviation from the average is no more than 0.3%. Figure 4 shows that time-averaged flow moves from wall to wall and is significantly asymmetrical.

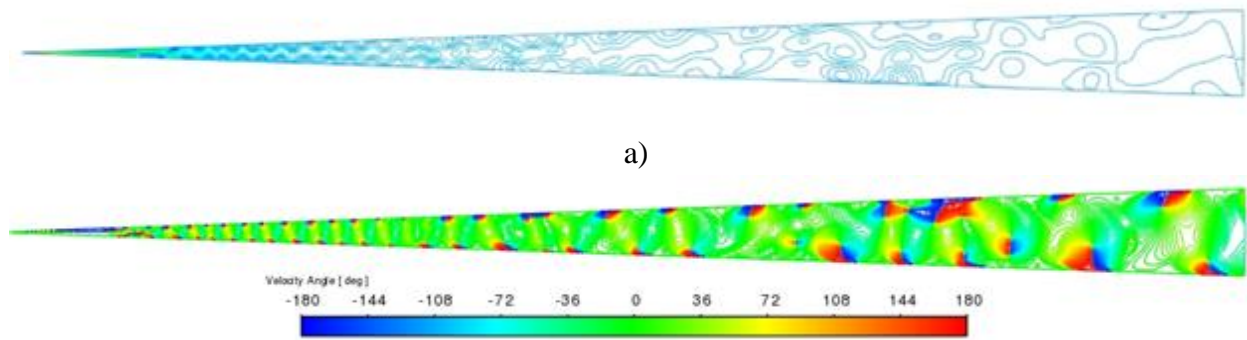

a)

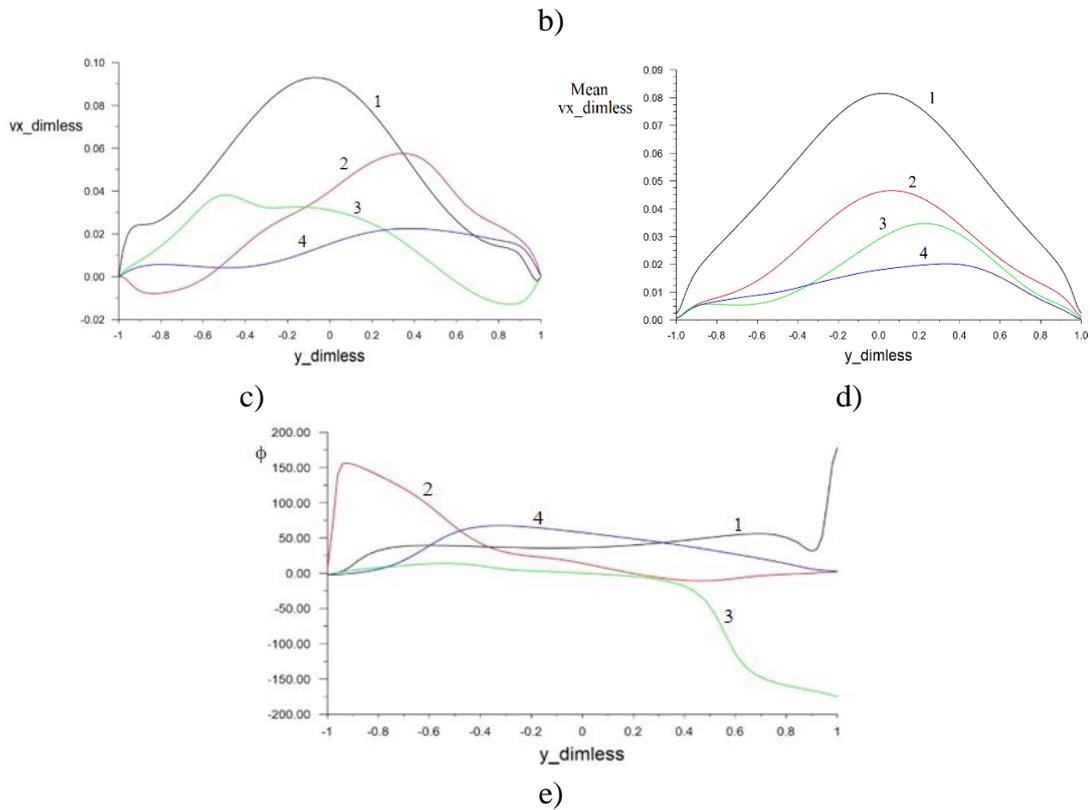

Fig. 4. The isolines of the horizontal component of average velocity mean_$V_x$ (a), isolines of the angles $\phi$ of velocity vectors (b) and profiles for dimensionless components of velocity (c, d) and profiles of the angles $\phi$ (e) for four vertical sections (lines 1 - 4 correspond to cross sections with $x_{dimless} = 20,\ 40,\ 60,\ 80$) at Reynolds number Re = 559.

### 3.2 Vibration effects from the walls of the diffuser

*Re=279, Re$_{vibr}$=0,02, Sh= 3.5*

In Fig. 5 the results of simulation the flow of a viscous liquid in a diffuser at steady state for Re=279 under vibration from the walls of the diffuser according to the law $V_n = A\sin(2\pi f t)$ (Re$_{vibr}$=0.02, Sh= 0.35) are presented. Figure 5 shows velocity isolines (a), isolines of the angles $\phi$ (b), velocity profiles (c) and profiles of the angles $\phi$ (d) in four vertical sections (lines 1 - 4 correspond to cross sections with $x_{dimless} = 20,\ 40,\ 60,\ 80$).

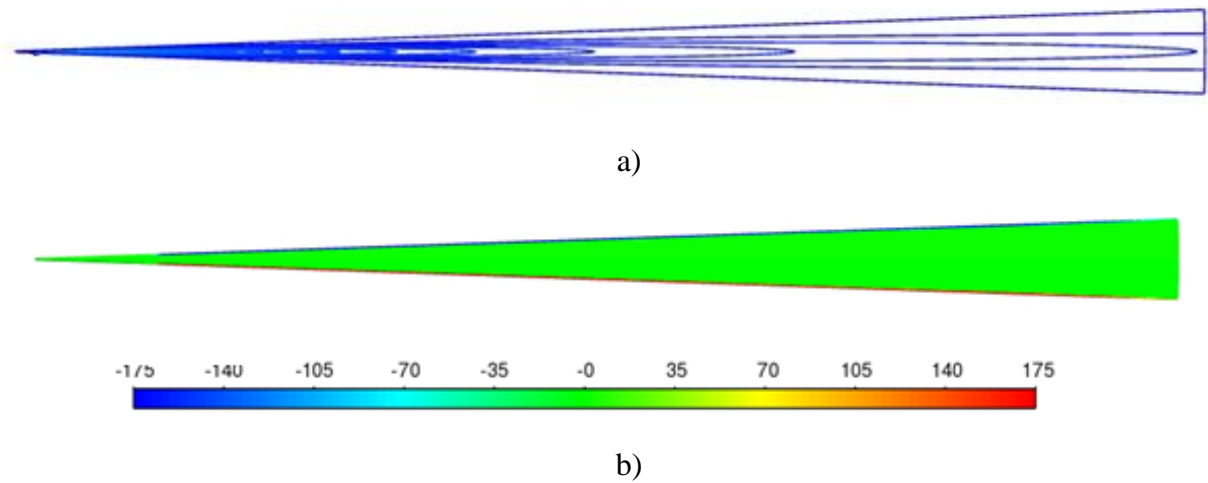

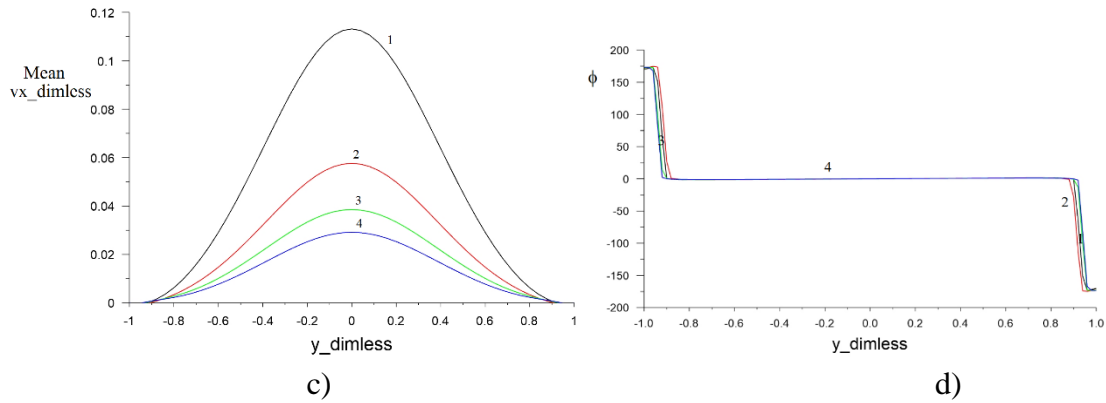

c)                                           d)

Fig. 5. The flow of a viscous fluid in the diffuser under vibration action from the walls of the diffuser according to the law $V_n = A\sin(2\pi ft)$ at Re=279, Re$_{vibr}$=0.02, Sh=3.5; a) – isolines of the longitudinal time-average velocity component $V_x$, b) – isolines of the angle $\phi$, c) – profiles of the longitudinal component of the average dimensionless velocities mean_$V_x$, and d) – profiles of the angles $\phi$ in vertical sections (lines 1 - 4 correspond to cross sections with $x_{dimless} = 20, 40, 60, 80$).

$$Re = 279,\ Re_{vibr} = 24,\ Sh = 3.5\ 10^{-3}$$

With intensive impact to the walls of the diffuser, when the homochronic number ($Sh^{-1}$) becomes commensurate with the Reynolds number Re, it is possible to influence not only the flow structure and velocity in the boundary layer near the wall, but also the shape of the main flow inside the diffuser. For such a case for $Re \cdot Sh \simeq 1$ ($Re = 279, Re_{vibr} = 24, Sh = 3.5\,10^{-3}$) in Figure 6 the following are shown: a) mean_Vx average velocity isolines, b) mean_vx average velocity profiles, and c) the profiles of angles $\phi$. Based on the results shown in Figure 6, we can see that intense vibrations from the walls of the diffuser generate an internal narrow jet along the entire length of the diffuser. The need for knowledge and the ability to create such flows is important for practical applications.

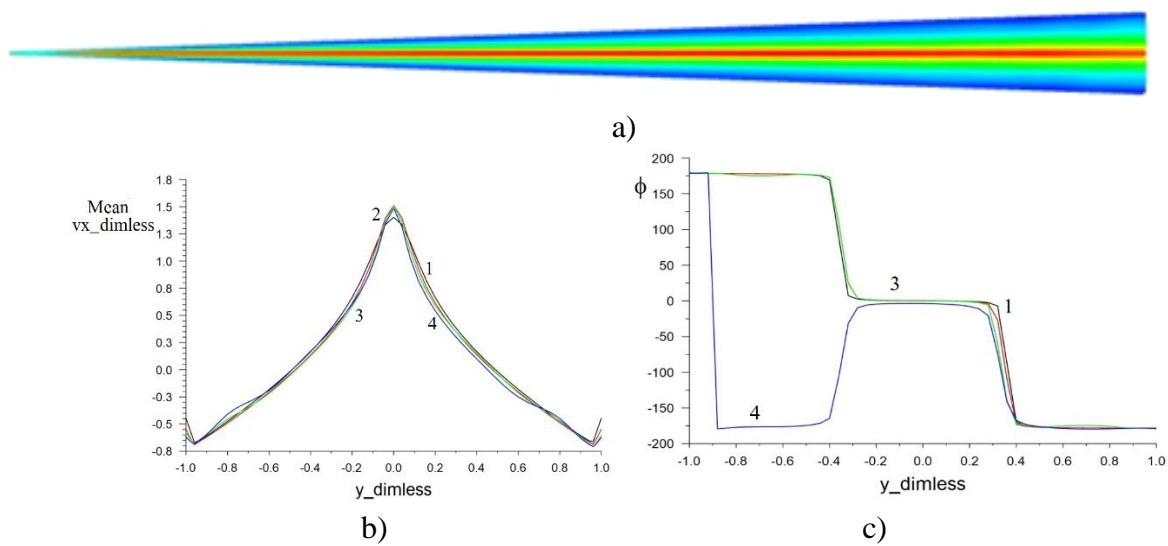

Fig. 6. Isolines of the longitudinal time-average velocity component $V_x$ (a), profiles of the longitudinal component of the average dimensionless velocity $V_x$ (a) and profiles of the angles $\phi$

(d) in vertical sections (lines 1-4, correspond to cross sections at $x_{dimless} = 20, 40, 60, 80$) under vibration action from the walls of the diffuser at $Re = 279$, $Re_{vibr} = 24$, $Sh = 3.5\cdot 10^{-3}$.

*Re=559, Re$_{vibr}$=2.4, Sh =0.35*

For the unsteady flow mode, Figure 7 shows the results for the Reynolds number Re=559 under vibration action from the walls of the diffuser for Re=559, $Re_{vibr}$=2.4, Sh=0.35. Figure 7 shows velocity isolines (a), isolines of the angles $\phi$, b), velocity profiles mean_Vx (c) and profiles of the angles $\phi$ (d) in four vertical sections (lines 1-4, correspond to cross sections at $x_{dimless} = 20, 40, 60, 80$).

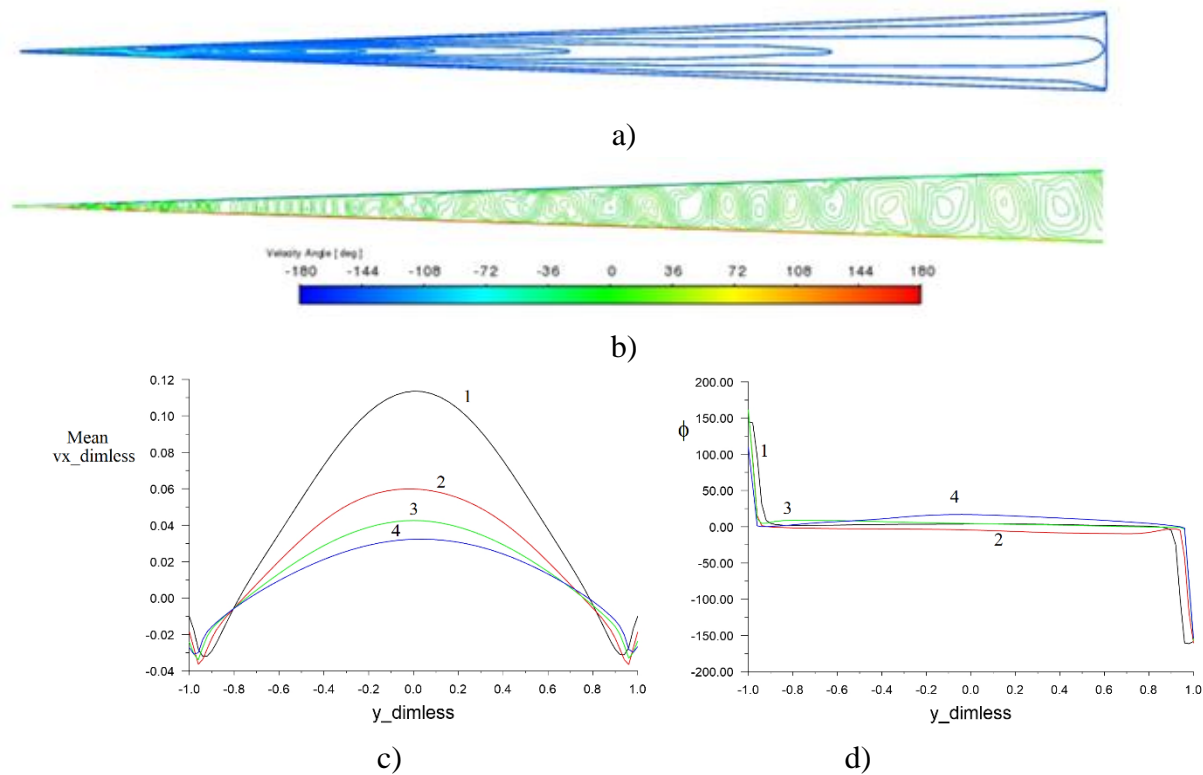

Fig. 7. Unsteady flow of a viscous liquid in a diffuser under vibration actions from the walls of the diffuser for Re=559, $Re_{vibr}$=2.4, Sh=0.35; a) – isolines of the longitudinal component of the average velocity mean_V$_x$ , b) – isolines of the angle $\phi$, c) – profiles of the averaged horizontal component of the velocity mean_V$_x$ and profiles of the angles $\phi$ in vertical sections (lines 1-4, correspond to cross sections at $x_{dimless} = 20, 40, 60, 80$)).

Comparing the flow results for Re=559 without vibration actions (Fig.4) and with vibrations (Fig.7), it can be seen that the action of low-intensity vibrations ($Re_{vibr}$=2.4, Sh=0.35) improves the symmetry of the unsteady fluid flow inside the diffuser at a Reynolds number of Re=559. However, this improvement does not occur throughout the entire length of the diffuser (Fig. 7a). Therefore, additional studies are needed to select the amplitude-frequency characteristics of these effects in order to achieve complete flow symmetry inside the diffusers at high Reynolds numbers. These studies may also include studying the effects of heterogeneous vibrations with different intensities along the diffuser length.

## 3.3 The vibration effects from the inlet to the diffuser

*Re=279, Re$_{vibr}$=2.4, Sh=3.5 10$^{-2}$*

It is known that the velocity and structure of the flow in a diffuser can be affected by vibrations applied at the inlet to the diffuser [10, 20, 21]. This article demonstrates the possibility of symmetrization of asymmetric flows at Re = 279, as shown in Fig. 3. Figure 8 shows the results demonstrating the effect of weak harmonic vibration ($Re_{vibr}$=2.4, Sh=3.5·10$^{-2}$) on the main flow in the diffuser at Re=279 where $Re_{vibr}$=2.4 corresponds to approximately 1% of velocity of main flow. A periodic vibration effect was applied to the velocity of the main flow at the inlet of the diffuser (towards the arc $l_{in}$ in a normal direction), in the form of a harmonic function: $V = V_{in} + A \sin(2\pi ft)$. Figure 8 shows the isolines of the instantaneous velocity $V_x$(a), the isolines of the mean velocity mean_$V_x$ (b), the isolines of the angles $\phi$ (c), the profiles of the dimensionless mean velocity mean_$V_x$ (d) and the profiles of the angles $\phi$ (e) in four vertical sections: $x_{dimless} = 20, 40, 60, 80$. In this case, it should be noted that the periodic vibration effects at the inlet ($Re_{vibr}$=2.4) symmetrizes the flow (Fig. 8b, d), but generates a weak wave structure along almost the length of the diffuser. This can be seen from the comparison of the fields of instantaneous (Fig. 8a) and averaged (Fig. 8b) velocities. Isolines (Fig. 8c) and angular profiles (Fig.8e) show that the flow is unidirectional along the entire length, except for small weak vortices located on the walls of the diffuser in a staggered manner (Fig 8c). Therefore, the results demonstrate that asymmetric flow can become symmetrical through vibration effects at the entrance to the diffuser. Results on vibration effects on flow in a diffuser for other parameters can be found in articles [20, 21].

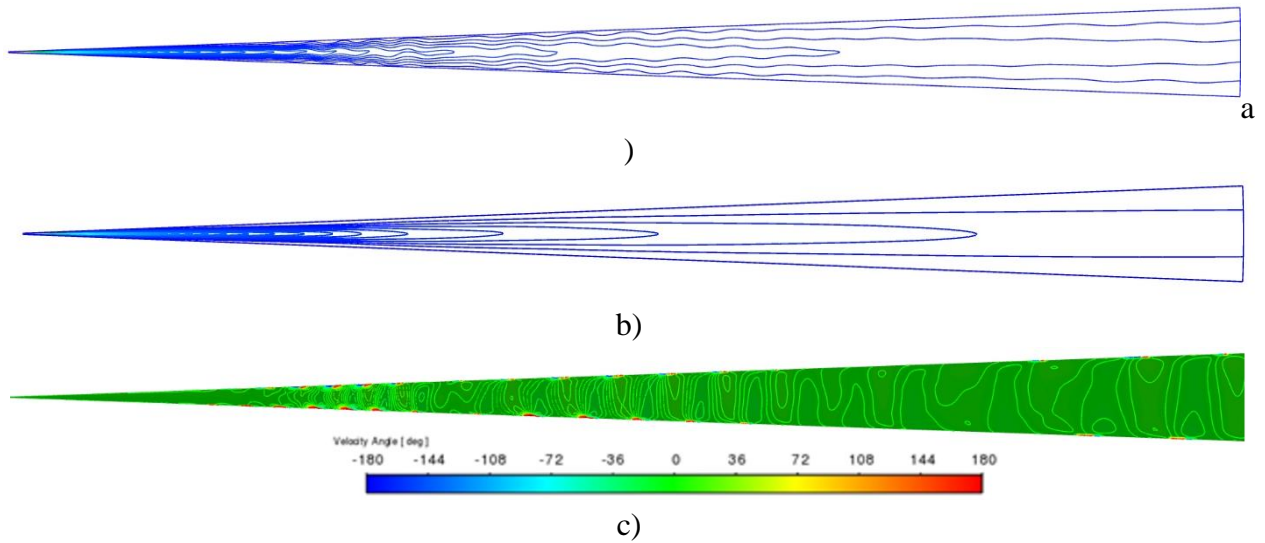

a)

b)

c)

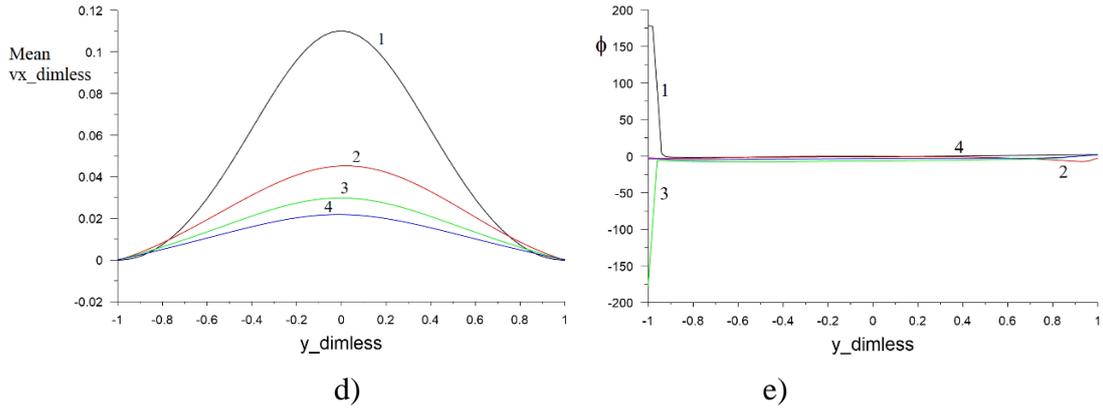

d)    e)

Fig. 8. Flow under vibration actions from the inlet boundary of the diffuser
$V = V_{in} + A \sin(2\pi ft)$ at Re=279, $Re_{vibr}$=24, Sh=3.5 $10^{-2}$; a) – isolines of the longitudinal component of the instantaneous velocity $V_x$; b) – isolines of the longitudinal component of the time–averaged velocity mean_$V_x$; c) – isolines of the angles $\phi$; d) – profiles of the averaged horizontal velocity component mean_$V_x$ and e) – profiles of the angles $\phi$ in vertical sections (lines 1-4 correspond to the sections $x_{dimless} = 20, 40, 60, 80$).

The results of parametric numerical calculations have shown that, of the two described methods of flow symmetrization, the most effective is the method with a vibrational effects on the boundary layer from the side of solid walls, rather than vibrational effects from inlet of diffuser. However, the method of influencing flow from entrance to the diffuser is simpler to implement, and is also an active method for symmetrizing of asymmetric flows in the diffusers.

### 3.4 The Richardson Effect
*Re=279, $Re_{vibr}$=24, Sh=3.5 $10^{-3}$*

When a periodic velocity (or pressure) is set at the entrance to the diffuser in form $V_{in} = A \sin(2\pi ft)$, a periodic constant flow occurs in the diffuser with a velocity profile different from the Poiseuille profile. Figure 9 shows the structure of the fluid flow with periodic velocity changes at the inlet to the diffuser in the form of $V_{in} = A \sin(2\pi ft)$ for the following dimensionless parameters $Re = 279$, $Re_{vibr} = 24$, $Sh = 3.5 \times 10^{-3}$. The profiles of the average longitudinal velocity component (Fig. 9) have velocity maxima near the walls – this is the Richardson effect [22], while the flow structure is symmetrical. This flow structure is like the Richardson ring effect, which occurs during the periodic flow of a viscous liquid in pipes [22]. Additional information can be found in [20,21].

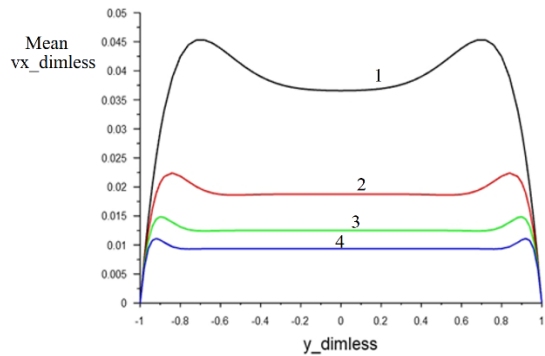

Fig. 9. Vertical profiles of the longitudinal component of the average dimensionless velocity mean Vx of the periodic liquid flow in the diffuser at
$Re = 279$, $Re_{vibr} = 24$, $Sh = 3.5 \times 10^{-3}$ (lines 1-4 correspond to cross sections with $x_{dimless} = 20, 40, 60, 80$).

## 4. CONCLUSIONS

The results of numerical modeling showed that asymmetric flows of viscous liquids can be symmetrized by vibrational effects on the velocity of the main flow at the inlet or on the walls of the diffuser. The low-amplitude vibration method is optimal for effectively symmetrizing the flow with minimal negative influence on the velocity of main flow. For example, to symmetrize a flow with Reynolds number Re=279 and opening angle $\beta = 4°$, vibration intensity can be less than 1% of maximum main flow velocity in the first case and less than 0.01% in the second case.

The simulation results showed that, when setting the periodic input velocity ($V_{inlet} = A \sin(2\pi ft)$) in a plane diffuser, there is an effect of Richardson with maxima near the walls in the transverse profiles of longitudinal velocity, which is known to occur for periodic fluid flow in pipes [22].

## FUNDING

This work was supported by the Russian Science Foundation grant 24–29-00101.

## CONFLICT OF INTEREST

The authors declare that they have no conflicts of interest.

## REFERENCES

1. Jeffery G.B. The two-dimensional steady motion of a viscous fluid. Phil. Mag.. Ser.6. **29**. (172), (1915) pp. 455–465.

2. Hamel G. Spiralformige Bewegungen zaher Flussigkeiten./ Jahres her. Deutsch. Math. Ver. Bd **25**, (1917) pp. 34–60.

3. Chang P. K. *Separation of Flow*. (Pergamon Press, London, 1970). 760 pp.

4. Schlichting H., Gersten K. *Boundary-Layer theory*. 9-th ed. (Springer Berlin, Heidelberg. 2017). 855p. DOI: https://doi.org/10.1007/978-3-662-52919-5.

5.     Rosenhead L. The steady two-dimensional radial flow of viscous fluid between two inclined plane walls. Proc Roy Soc London Ser A;**175**(963) (1940) 436–67.

6.     Fraenkel L. E. Laminar flow in symmetrical channels with slightly curved walls. I. On the Jeffery–Hamel Solutions for flow between plane walls. Proc. R. Soc. Lond. Ser. A, Math.Phys. Sci. **267** (1328) (1962), pp.119–138.

7.     Goldshtik M., Hussain F. and Shtern V. Symmetry breaking in vortex-source and Jeffery-Hamel flows // J. Fluid Mech., **232**, (1991) pp. 521-566.

8.     Ashjaee, J., Johnston, J. P. Straight walled two-dimensional diffusers -transitory stall and peak pressure recovery. J. Fluids Eng. **102**: (1980) pp. 275- 282.

9.     Durst F., Melling A., Whitelaw J.H. Low Reynolds number flow over a plane symmetric sudden expansion. Journal of Fluid Mechanics, **64**, (01), (June 1974), pp 111-128.

10.    Nabavi M. Three-dimensional asymmetric flow through a planar diffuser: Effects of divergence angle, Reynolds number and aspect ratio. International Communications in Heat and Mass Transfer **37**, (2010), pp. 17–20

11.    Pukhnachev V.V. Symmetry in the Navier–Stokes equations (Uspekhi Mekhaniki, 2006), pp. 6–76. (In Russian).

12.    Fedyushkin A.I. The Transition Flows of a Viscous Incompressible Fluid in a Plane Diffuser from Symmetric to Asymmetric and to Non-Stationary Regimes. Physical-Chemical Kinetics in Gas Dynamics. **17** (3) (2016). http://chemphys.edu.ru/issues/2016-17-3/articles/638/

13.    Kerswell, R. R., Tutty, O. R. & Drazin, P. G. Steady nonlinear waves in diverging channel flow. J. Fluid Mech. **501**, (2004) 231–250.

14.    Dennis S. C. R., Banks W. H. H., Drazin P. G. and Zaturska M. B. Flow along a diverging channel. J. Fluid Mech. (1997. **336**, pp. 183-202.

15.    Haines P.E., Hewitt R.E., and Hazel A.L. The Jeffery – Hamel similarity solution and its relation to flow in a diverging channel. J. Fluid Mech., (2011). **687**. pp. 404–430.

16.    Bhoraniya, R., Swaminathan, G., Narayanan, V. Diverging Channel. In: Global Stability Analysis of Shear Flows. Springer Tracts in Mechanical Engineering. (Springer, Singapore. 2023). https://doi.org/10.1007/978-981-19-9574-3_3

17.    Akulenko, L.D., Georgievskii, D.V., Kumakshev, S.A.: Solutions of the Jeffery-Hamel problem regularly extendable in the Reynolds number. Fluid Dyn. (2004). **39**(1), 12–28.

18.    Akulenko L.D., Kumakshev S.A. Bifurcation of multimode flows of a viscous fluid in a plane diverging channel // Journal of Applied Mathematics and Mechanics. (2008). **72**. pp 296–302.

19.    Fedyushkin A.I., Volkov E.V. The Symmetry of the Flow of Newtonian and Non-Newtonian Fluid in a Plane Diffuser and Confusor. Physical-Chemical Kinetics in Gas Dynamics (2019) **20** (2). http://chemphys.edu.ru/issues/2019-20-2/articles/791/

20.    Fedyushkin A.I. The Influence of Controlled Vibration Effects on Fluid Flow in Technological and Engineering Processes. In: Feng, G. (eds) Proceedings of the 10th International Conference on Civil Engineering. ICCE 2023. Lecture Notes in Civil Engineering, vol 526. Springer, Singapore. (2024). https://doi.org/10.1007/978-981-97-4355-1_64.

21. Fedyushkin A., Puntus A. Symmetrisation of laminar flow of viscous fluid in a flat diffuser by periodic influence on the inlet flow velocity // E3S Web of Conferences. (2023). Vol. 446. 01001.

22. Richardson E.G. and Tyler E. The transverse velocity gradient near the mouths of pipes in which an alternating or continuous flow of air is established. Pros Phys Soc London. **42**, (1). (1929). 7–14. DOI: 10.1088/0959-5309/42/1/302

23. Patankar S. V. *Numerical Heat Transfer and Fluid Flow*. (Hemisphere. 1980). 214p.

24. Polezhaev V I, Bello M S, Verezub N A et al. *Convective Processes in Weightlessness* (M: «Nauka», 1991).240p. (in Russian).

25. Polezhaev V. I., Bune A. V., Verezub N A, et al. *Mathematical modeling of convective heat and mass transfer based on Navier-Stokes equations*. (M.: «Nauka», 1987). 272p. (in Russian).

26. Fedyushkin A. I. "Research of a matrix method for solving convection equations. Complex of programs MARENA" Preprint 471 M.: IPM of the USSR Academy of Sciences 32p. (1990). (in Russian).

27. Fedyushkin, A.I., Ivanov, K.A. & Puntus, A.A. "An Effective Multigrid Method for Solving Problems of High-Frequency Vibrational Convection". J. Appl. Ind. Math. **17**. 307–319. (2023). https://doi.org/10.1134/S1990478923020096